\documentclass[prd,aps,twocolumn,nofootinbib]{revtex4}
\usepackage{graphicx}
\usepackage{amssymb}
\usepackage{color}
\def\gr{$\gamma$-ray}
\def\an{\textcolor{black}}
\begin{document}

\title{Origin of TeV Galactic Cosmic Rays}
\author{A.Neronov$^1$} \email{Andrii.Neronov@unige.ch}  \author{D.V.Semikoz$^{2,3}$} \email{dmitri.semikoz@apc.univ-paris7.fr}
   \affiliation{$^1$ ISDC Data Centre for Astrophysics, Ch. d'Ecogia 16, 1290, Versoix, Switzerland}
                     
             \affiliation{$^2$ APC, 10 rue Alice Domon et Leonie Duquet, F-75205 Paris Cedex 13, France}
\affiliation{$^3$ Institute for Nuclear Research RAS, 60th October Anniversary prosp. 7a, Moscow, 117312, Russia}

\begin{abstract}
We consider a possibility of identification of sources of cosmic rays (CR) of the energy above 1~TeV via observation of degree-scale extended \gr\ emission which traces the locations of recent sources in the Galaxy.   Such emission in the energy band above 100~GeV is produced by CR nuclei and electrons released by the sources and spreading into the interstellar medium.  We use the data from the Fermi \gr\ telescope to locate the degree-scale 100~GeV \gr\  sources. We find that the number of such sources and their overall power match to those expected when CRs injection events happen every $\sim 100$~yr in portions of $\sim 10^{50}$~erg. We find that most of the sources are associated to pulsars with spin down age less than $\sim 30$~kyr and hence to the recent supernova explosions. This supports the hypothesis of supernova origin of Galactic CRs. We notice that the degree-scale extended emission does not surround shell-like supernova remnants without pulsars. Based on this observation,  we argue that the presence of the pulsar is essential for the CR acceleration process. We expect that a significant fraction of the degree-scale sources should be detectable as extended sources with km$^3$-scale neutrino detectors. 
\end{abstract}
\maketitle

\section{Introduction}

The bulk of the Cosmic Ray (CR) flux hitting the Earth's atmosphere is composed of protons and heavy nuclei with energies in the 1-10~GeV range \cite{ginzburg,berezinski}. They are injected into the interstellar medium (ISM) by unknown sources and spread across the ISM via diffusion in the turbulent interstellar magnetic field. The diffusive  propagation of CRs provides a major obstacle for identification of the CR sources: angular distribution and spectral properties of CRs captured by particle detectors mostly provide information on the details of the process of diffusion of CRs through the Galactic magnetic field rather than on the location of the CR sources and mechanisms of particle acceleration in the sources. 

Information on the details of diffusion of the CRs through the ISM, obtained from the particle detector measurements, provides a constraint on the energy output of the Galactic CR sources. The CR data suggest that typical CR particles spend only $t_{CR}\le10^7$~yr in the Galactic Disk before escaping to the larger Halo and/or to the intergalactic medium \cite{ginzburg,berezinski,strong10,blasi}. Constraint on the lifetime of CRs in the Galaxy imposes a constraint on the cumulative time-average power of the CR sources, $L_{CR}\sim 10^{41}$~erg/s \cite{ginzburg,berezinski,strong10,blasi}, needed to support the observed CR energy density $\simeq 1$~eV/cm$^3$ \cite{prl}. 

This power could be supplied e.g. by  supernovae if some $\sim 0.1\%$ of the energy output of the gravitational collapse of the star is  transferred to the CRs \cite{zwicky34,ginzburg}. Several types of the supernova-related phenomena are known to be associated to particle acceleration: \gr\ bursts (GRB), pulsars and the associated pulsar wind nebulae (PWN) and  supernova remnant (SNR) shells.  Particle acceleration to the energies $\gtrsim 1-10$~GeV is now firmly established from the \gr\ band observations of all the three types of objects \cite{review,review_fermi}.

Alternatively, the required power could be provided by acceleration process which operates at the distance and time scales scales larger than individual supernova remnants, e.g. in superbubbles \cite{bykov,parizot,ackermann} and/or even continuously throughout the Galaxy (Ref. \cite{butt} and references therein). Evidence in favor of such scenario is found from the study of chemical composition of the low-energy cosmic ray flux \cite{lingenfelter07,binns08,wiedebeck99} and from the direct association of GeV band extended emission with particle acceleration in Cygnus superbubble \cite{ackermann}.

Clarification of the physical mechanism responsible for the CR production requires localization of the CR sources in space.  This could, in principle, be done using \gr\ observations. This possibility arises because CRs confined in the source could interact with low-energy particles  producing \gr s and neutrinos. However, identification of the CR sources among the Galactic \gr\ sources suffers from large uncertainties. In most of the cases it is not clear if the amount of target material in the source is sufficiently large for the  production of detectable CR interaction related \gr\ emission. Moreover, in most of the sources protons and nuclei are accelerated together with electrons.  The accelerated electrons produce \gr\ emission via inverse Compton and Bremsstrahlung mechanisms, with intensity larger than that of the \gr\ emission from CR interactions. As a result, the "hadronic" component of the \gr\ flux is hidden behind much stronger "leptonic" \gr\ emission \cite{aharonian_book}. If CRs are injected by a process operating on large distance scales in the Galaxy, \gr\ observations might not reveal individual isolated CR sources at all \cite{butt}. 

Taking into account these difficulties, we consider in this paper an alternative method for the identification of the Galactic CR sources based on the \gr\ data. Instead of searching for the \gr\ emission from CR interactions in the sources, we consider the \gr\ emission produced by CRs which escape from the source and start to spread into the ISM. CRs penetrating into the ISM interact with the ISM particles and produce \gr s  via the same mechanism as inside the sources, but on much larger distance and time scales. The overabundance of CRs around the individual sources leads to the increased CR -- ISM interaction rate from an extended region around the source \cite{aharonian_book}. This results in the appearance of  extended \gr\ emission.  We show that such extended emission around CR sources becomes detectable in the very-high-energy (VHE) band above $\sim 100$~GeV. We identify the extended Galactic sources of $\gtrsim 100$~GeV \gr s in the data of the Large Area Telescope (LAT) on board of Fermi satellite \cite{atwood09}. We study  the overall distribution and energetics of the degree-scale sources in the Galaxy. The source statistics and morphology  are consistent with the assumption that the extended \gr\ emission is produced by the CR interactions. We show that extended VHE \gr\ emission  is found around all known nearby pulsar-producing supernova events which occurred in the last 30~kyr, as expected if the Galactic CRs are injected by the supernovae. We further argue that the presence of the young pulsars inside most of the degree-scale extended sources points to the possibility that CRs with energies above 1~TeV are injected by the PWN or by the composite SNR (shell-like SNR confining a PWN) systems, while distributed acceleration of cosmic rays in shocks produced by multiple supernovae and stellar wind bubbles might be most important at lower energies.

\section{Locating the CR sources}
\label{sec:locating}

 CRs which escape from  sources can  be traced  as they spread  into the ISM. They could be detected via the \gr\ emission produced in CR interactions with the ISM particles.  CRs of the energy $E_{CR}$ diffuse through the Galactic magnetic field. Measurements of the energy-dependent diffusion coefficient based on the local measurements of the primary and secondary CR nuclei and on the modeling of the diffuse \gr\ emission from the Galaxy  suggest the value
\begin{equation}
\label{D}
D  =  D_{28} \times 10^{28}\left[ E_{CR}/4\mbox{ GeV}\right]^{-\delta}~\mbox{cm}^2\mbox{/s}, 
\end{equation}
with an uncertainty of the prefactor $D_{28}$ by up to a factor of $\simeq 3$ and with the slope $\delta=0.4\pm 0.1$ \cite{strong10}. CRs released from a source some $T_s$ years ago are now filling a volume of  size \cite{blasi}
\begin{equation}
\label{rs}
r_s\simeq 2\sqrt{DT_s}\simeq 80\ D_{28}^{1/2}\left[\frac{T_s}{10\mbox{ kyr}}\right]^{1/2}\left[\frac{E_{CR}}{1\mbox{ TeV}}\right]^{\delta/2}\mbox{ pc.}
\end{equation}
The average density of ISM on this distance scale is $n_{ISM}\sim 1$~cm$^3$  \cite{korchagin03}. The typical interaction time of CRs propagating through the ISM is 
\begin{equation}
\label{tpp}
t_{pp}=\left(c\sigma_{pp}n_{ISM}\right)^{-1}\simeq 3\times 10^7\left[\frac{n_{ISM}}{1\mbox{ cm}^{-3}}\right]^{-1}\mbox{ yr,}
\end{equation}
where we adopt an estimate $\sigma_{pp}\simeq 4\times 10^{-26}$~cm$^2$ for the cross-section of inelastic proton-proton collisions for the TeV CRs \cite{pdg}. 

CR interactions with ISM inside a region of the size $r_s$  lead to extended  \gr\ (and neutrino) emission from the entire CR filled volume. The \gr\ luminosity  can be readily estimated taking into account that 
each source must release some  $E_s\sim 3\times 10^{50}\left[{\cal R}_{SN}/10^{-2}\mbox{ yr}\right]^{-1}$~erg in the form of CRs, where ${\cal R}_{SN}$ is the rate of CR injection events in the Galaxy.  This leads to the estimate
\begin{equation}
\label{lgamma}
L_\gamma\sim \frac{\kappa  E_{s}}{t_{pp}}\sim 2\times 10^{34}\left[\frac{\kappa}{0.2}\right]\left[\frac{E_s}{10^{50}\mbox{ erg}}\right]\left[\frac{n_{ISM}}{1\mbox{ cm}^{-3}}\right]\mbox{ erg/s,}
\end{equation}
where $\kappa\sim 0.2$ is a typical fraction of the CR energy deposited in the \gr s. Note that this luminosity depends only on the average density of the ISM. It is largely independent of the physical parameters of the source itself. The only source parameter which is fixed by the general energy requirement on the Galactic CR sources is the released energy in the form of CRs, $E_s$. The estimate of $L_\gamma$ in Eq. (\ref{lgamma}) is uncertain by at least a factor of $\sim 3$ \cite{ginzburg,berezinski}. This includes the uncertainty of the estimate of the typical energy output per source, $E_s$, which depends on the lifetime of CRs in the Galactic disk, the thickness of the disk and the rate of the injection events (with large uncertainties in all the three factors). Besides, the estimate of $L_\gamma$ is energy dependent, due to the energy dependence of the source power and of the CR lifetime in the Galactic Disk.  

\section{Extended VHE \gr\ emission around CR sources}
\label{sec:extended}

The luminosity level given in Eq. (\ref{lgamma}) is high enough for the \gr\ emission to be detectable with existing space- and ground-based \gr\ telescopes. The main difficulty for the detection of extended emission around the individual CR sources is that it has to be identified on top of the diffuse \gr\ emission from the Galaxy \cite{abdo09}. 

 In the energy band 0.1-1~GeV this diffuse emission appears as a collective emission from a very large number of extended sources. Possible emission around individual sources  can not be identified. The difficulty for the  identification of individual sources is, in fact, the problem of the source confusion. The 0.1-1~GeV \gr s are produced by the CRs with energies in the $E_{CR}\sim 1-10$~GeV range. In this energy range the size of the extended emission region around the source reaches the $\gtrsim 100$~pc scale in  $\sim 10^5$~yr (see Eq. (\ref{rs})). Assuming the supernova rate of ${\cal R}_{SN}\sim 10^{-2}$~yr$^{-1}$, one finds that the 100~pc scale regions around some $N_s\sim 10^3/D_{28}\left({\cal R}_{SN}/10^{-2}\mbox{ yr}^{-1}\right)$ sources are currently filling the Galactic Disk. The density of the sources is $\rho_s\sim N_s/V_{disk}\sim 3$~kpc$^{-3}$, where $V\simeq 2\pi (10\mbox{ kpc})^2H_{disk}\sim 300$~kpc$^3$ is the volume of the Galactic Disk. This implies typical distance between the centers of the extended emission, $\sim 700$~pc, so that in any direction in the inner Galaxy one would find on average several extended sources  within several kiloparsec distance are superimposed on each other. An additional difficulty is the large size of the point-spread function (PSF) of the LAT in this energy band \cite{atwood09}.

The number of sources surrounded by the extended emission decreases with the increasing energy. At energies $E_\gamma\gtrsim 100$~GeV (the CR energies in the TeV range), the extended emission region size reaches the 100~pc  in $\sim 10$~kyr (see Eq. (\ref{rs})). Some $\lesssim 100$ sources are currently surrounded by the 100~pc scale extended emission, so  that the density of the sources decreases down to $\sim 0.3$~kpc$^{-3}$. The typical source-to-source distance is $\sim 1.5$~kpc so that there is no overlap.  Relatively compact and bright extended \gr\ emission  around nearby and recent CR sources should start to be identifiable in the data. 

A source which released CRs some $T_s$~yr ago is surrounded by extended emission of the  angular size
\begin{equation}
\theta_s\sim \frac{r_s}{R_s}\simeq 0.8^\circ D_{28}^{1/2}\left[\frac{R_s}{5\mbox{ kpc}}\right]^{-1}\left[\frac{T_s}{10\mbox{ kyr}}\right]^{1/2}
\left[\frac{E_{CR}}{1\mbox{ TeV}}\right]^{0.2},
\label{thetas}
\end{equation}
where $R_s$ is the distance to the source. The expected  \gr\ flux is  
\begin{equation}
\label{fgamma}
F_s=\frac{L_\gamma}{4\pi R_s^2}\simeq  10^{-11}\left[\frac{R_s}{5\mbox{ kpc}}\right]^{-2}\left[\frac{n_{ISM}}{1\mbox{ cm}^{-3}}\right]\left[\frac{\kappa}{0.2}\right]\frac{\mbox{ erg}}{\mbox{cm}^2\mbox{s}}.
\end{equation}
A reasonable task is to search for the relatively compact \gr\ emission around the TeV CR sources of the age $T_s\sim 10^4$~yr on top of the diffuse background.  Up to
\begin{equation}
N_s\sim \rho_s\pi R_s^2/2\sim 12D_{28}^{-1}\left[\frac{R_s}{5\mbox{ kpc}}\right]^2\left[\frac{T_s}{10\mbox{ kyr}}\right]^{1/2}\left[\frac{{\cal R}_{SN}}{1/100\mbox{ yr}}\right]
\label{ns}
\end{equation}
degree-scale extended sources could be detectable within the $\sim 5$~kpc distance in the inner Galaxy where the density of the ISM is high enough to provide the required luminosity level. 

Detection / non-detection of the degree-scale sources at the flux level given by Eq. \ref{fgamma} provides a test for the supernova scenario of CR production. Indeed, non-detection  would rule out the possibility that the CRs are injected in portions of $\sim 10^{50}$~erg once every $\sim 100$~yr and would instead point to a scenario where the CRs are injected continuously throughout the Galaxy, as it is expected in the scenario of cosmic ray production via acceleration at multiple shocks formed by the stellar wind bubbles and supernova explosions in the star forming regions. It is, however, worth mentioning that in this scenario the degree-scale extended emission might still originate from the individual most active superbubbles which are currently accelerating cosmic rays.

\section{Search for the degree-scale sources in the LAT data}
\label{sec:bubbles}

The results obtained in the previous section suggest that the nearby and relatively recent sources of TeV CRs should reveal themselves though the degree-scale extended \gr\ emission. The expected flux level (\ref{fgamma}) is high enough so that the LAT sensitivity at 100~GeV is sufficient for the source detection.  

We have analyzed the data collected in the period between August 2008 and October 2011 using the Fermi Science Tools\footnote{http://fermi.gsfc.nasa.gov/ssc/data/analysis/}. The LAT event lists were filtered using {\it gtselect} tool and only "superclean" \gr\ events ({\tt evtclass=4}) with energies above 100~GeV arriving at zenith angles $\theta_z<100^\circ$ were retained for the analysis. 
To identify the extended sources of the size $\theta_s\lesssim 1^\circ$ in the LAT data above 100~GeV we have used the Minimal Spanning Tree (MST) method~\cite{Canipana:2008zz,massaro,Neronov:2010qp}.  In this method photons separated by angular distances smaller 
then a pre-defined value $\theta_{max}$  are considered as belonging to a "cluster" or "tree". Point-like or extended sources could be found on top of the diffuse background as trees with large enough number of photons. Significance of the source detection can be estimated via comparison of the real data with Monte-Carlo (MC) simulations.

To increase the sensitivity of the MST method for the search of extended \gr\ sources we took $\theta_{max}$ in the range $0.3^\circ<\theta_{max}<0.5^\circ$, which is larger than the size of the PSF of the LAT telescope at the energies above 100~GeV.  In this respect the MST procedure is different from that used in the Ref. \cite{Neronov:2010qp}.  Close to the Galactic Plane, the level of diffuse sky background depends on the source location. The increase of $\theta_{max}$ leads to the increase of the chance coincidence probability of association of diffuse background photons with a tree.  To estimate the background at each source location we used MC simulations which generate the background maps from the real data by changing the sky photon coordinates $(l,b)$. In these MC  simulations the Galactic latitude of the photons was left unchanged, while the longitude was randomly redistributed within  $\Delta l=\pm 10^\circ$ around the real photon position.

\begin{figure}
\includegraphics[height=\columnwidth,angle=-90]{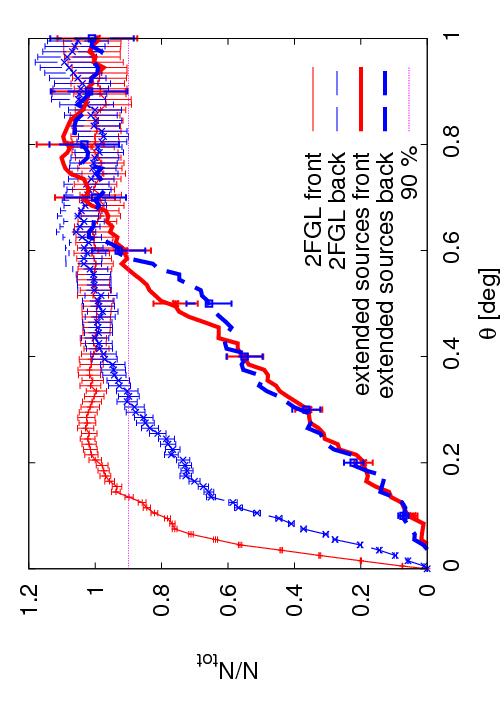}
\caption{Cumulative photon distribution as function of angle from the source position in the energy band above 100~GeV. Thin lines are for PSF around extragalactic Fermi sources. Thick lines are photon distribution for the Galactic sources from Table \ref{tab:list}. Solid line is for front photons, dashed line is for back photons.  Errorbars indicate the uncertainty due to background fluctuations. }
\label{fig:psf}
\end{figure}

We measure the extension of the sources by comparing the photon distribution around the source position with the PSF. To estimate the LAT PSF at $E>100$ GeV we took 1319 sources from second Fermi catalog \citep{fermi_catalog} with $|b|>10^\circ$ and calculated flux around them as function of angle. Results of this calculation are shown in the Fig.~\ref{fig:psf} separately for front and back photons. The 90\% containment radius is $\theta_{90,f} =0.14^\circ$ for front photons and  $\theta_{90,b} =0.33^\circ$ for back photons.  We consider the sources for which 90\% of $E>100$~GeV events are within $\theta_{90,f},\theta_{90,b}$ as point sources, while other sources are considered as extended. We remove photons associated to the point sources in the search for event clusters associated to extended sources. 

To identify point sources, we search for the correlation of the photon arrival directions with the positions of the sources from the two-year Fermi catalogue \citep{fermi_catalog}, in a way similar to the one used in the Ref. \citep{extragalactic} for the search of the extragalactic VHE \gr\ sources. The list of point sources found above $3\sigma$ significance threshold  is presented in the Table \ref{tab:ps}.  In this table we give the name of source in 2FGL catalog, galactic coordinates $l$ and $b$, the number of photons within  90\% C.L. angle from source, it's significance calculated using binomial probability, the type of the source and its alternative name for the identified sources.  We also looked for the point sources with MST method with photon separation $\theta=0.2^\circ$. This method does not rely on the pre-defined source catalogue. The search with the MST method did not result in additional point  sources which are not associated to the known 2FGL source.  

Ten out  of 14  point  sources listed in Tables \ref{tab:ps} are known VHE \gr\ sources. Four sources are new detections, VCS J0110+5805, 2FGL J1030.4-6015, PSR J1124-5916 and AT20G J160350-49. Five out of 14 sources are extragalactic sources found at low Galactic latitudes. Three sources are of unknown type. Other sources are known brightest Galactic \gr\ sources.  All the photons associated to the point sources listed in Table \ref{tab:ps} are subtracted from the overall photon list used in the search of extended sources. 

The list of extended sources found using the MST method is given in Table  \ref{tab:list}.  To reduce the amount of false trees formed by fluctuations of the background, we considered only trees with known counterparts from the TeVCat catalogue of VHE \gr\ sources\footnote{http://tevcat.uchicago.edu} for the source in the inner Galaxy ($-60^\circ<l<60^\circ$). After imposing such a requirement, two out of 14 extended sources in the inner Galaxy are expected to be due to background fluctuations.  

\begin{figure*}
\includegraphics[height=0.8\linewidth,angle=0]{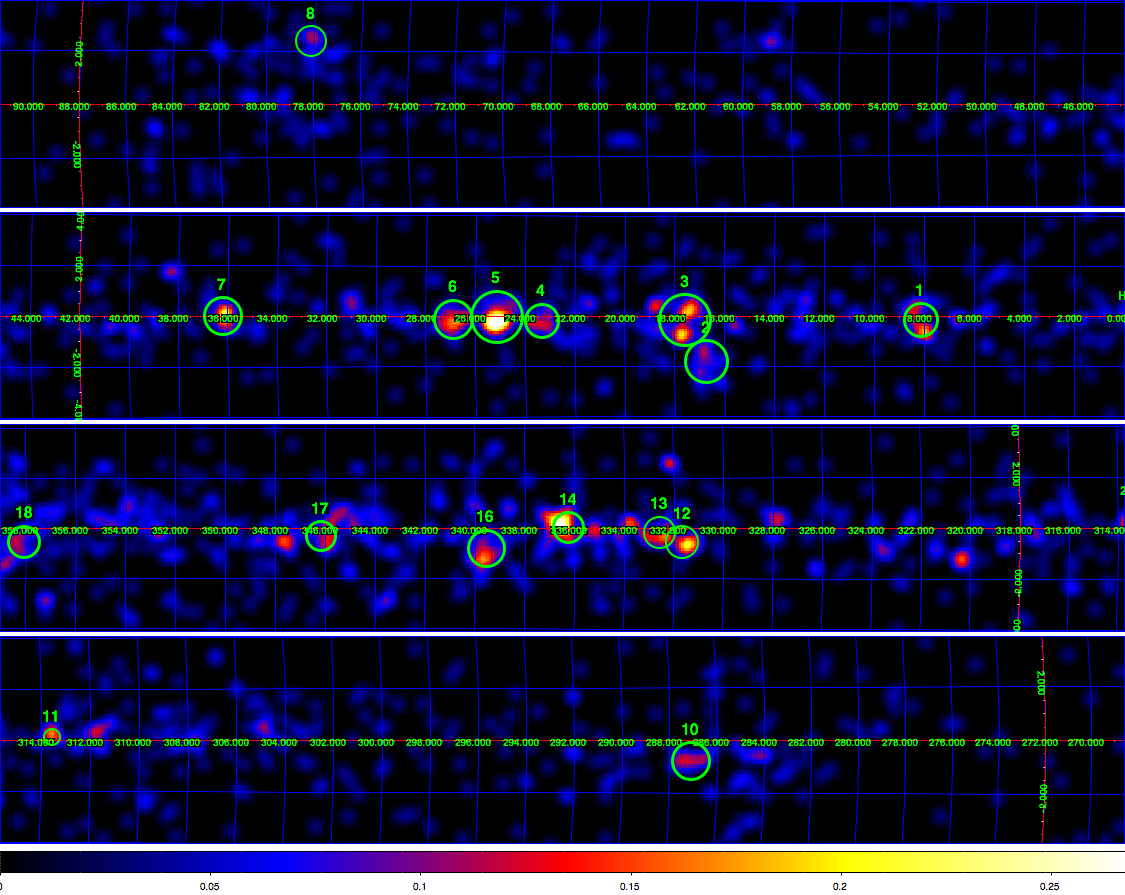}
\caption{LAT countmap of the inner Galaxy in the 100-400~GeV band smoothed with a 0.5~degree Gaussian. VHE \gr\ sources from Table  \ref{tab:list} are shown by green circles.} 
\label{fig:sky_map}
\end{figure*}

Fig. \ref{fig:sky_map} shows the LAT countmap in the energy range 0.1-0.4~TeV smoothed with 0.5~degree Gaussian. The positions of the sources listed in Tables  \ref{tab:ps} and  \ref{tab:list} are shown on the map.

\begin{table*}
\begin{tabular}{|c|c|c|c|c|c|c|c|}
\hline
 &2FGL                 &        $l$ & $b$       &$N_{ph}$&$P$  &    type &Name\\
 \hline
1&    1837.3-0700c     &25.09  &-0.08    &4  &1.e-4&& {\bf HESS J1837-069}\\
\hline
2&J2001.1+4352&    79.06 & -7.12     & 2             &1.e-3&BLZ  &  {\bf MAGIC J2001+435}					   \\	
\hline
3& J2323.4+5849&  111.74&  -2.11    & 2 &1.e-3&SNR&  {\bf Cas A}					   \\	
\hline
4 &J2347.0+5142&   112.88 & -9.90   & 4 &6.e-8& BLZ    & {\bf 1ES 2344+514}  			   \\
\hline
5&J0035.8+5951&     120.97 & -2.96   & 5 &4.e-8& BLZ   & {\bf 1ES 0033+595}				   \\
\hline
6&J0110.3+6805&     124.70 & 5.29   & 2 &6.e-4&   &VCS J0110+6805				   \\
\hline
7& J0240.5+6113&    135.67 & 1.08       & 4 &2.e-6 &GRLB& {\bf LS I+61 303}				   \\
\hline
8&J0521.7+2113 &   183.6& -8.70         & 4 & 2.e-5&AGU& {\bf VCS J0521+2112} 		   \\
\hline
9& J0534.5+2201&    184.55   &-5.78   & 28 & 0&PWN& {\bf Crab} 					   \\
  \hline
10& J0617.2+2234e&    189.05  &3.03     &  4& 7.e-5&SNR+CCO& {\bf IC443}				   \\
\hline
11& J0648.9+1516&    198.99   &6.35   &4 & 4.e-7&AGU&  {\bf VER J0648+152}					   \\
 \hline
12&  J1030.4-6015&     286.28  &-2.03    & 2&  1.e-3&    & 			   \\
 \hline
13&   J1124.6-5913&     292.2  &-2.03    & 2&  1.e-3     &PWN& PSR J1124-5916 				   \\
\hline
14&     J1603.8-4904      &332.15  &2.56      &5  &5.e-7  &   &AT20G J160350-49\\
\hline
\end{tabular}
\caption{List of point like VHE sources found  at the positions of known Fermi sources. $l,b$ are source coordinates, $N_{ph}$ is the number of photons within the 90\% containment radius of the PSF, $P$ is the chance coincidence probability for the event cluster to be a background fluctuation. Notations in the column "Type" are PWN: pulsar wind nebula; SNR+CCO: supernova remnant with compact central object, BLZ: blazar; AGU: AGN of unknown type; GRLB: \gr\ loud binary. Previously reported VHE \gr\ sources are marked in bold. Other sources are new detections in the VHE band. }
\label{tab:ps}
\end{table*}

\begin{table*}
\begin{tabular}{|l|l|l|l|l|l|l|l|l|l|l|l|l|}
\hline
& $l$ & $b$ &$\theta_{50}$ &$\theta_{90}$ &$P_{90}$&$N_{ph}$&$F$&Comments&SNR&PSR&$R_s$&$T_s$\\
\hline
1 &8.15 & -0.14 & 0.47 & 0.65 & 1.e-5 & 12 &$4.6\pm 1.3$&  {\bf HESS 1804-216}&W30&B1800-21 & 3.9 &1.6	   \\
 \hline
2 &16.74 & -1.79 & 0.46 & 0.83 & 1.e-6& 12 & &  {\bf LS  5039}  &&& &  				   \\
\hline
3 &17.58 & -0.14 & 0.6 & 1. 0& 1.e-3& 13 &$5.2\pm 1.4$&  {\bf HESS J1825-137}&&B1823-13& 4.1 & 2.1				   \\
 \hline
4 &23.32 & -0.16 &0.5  & 0.6 & 2.e-2& 8 &$3.4\pm 1.2$&  {\bf HESS J1834-087}	&W41&   CXOU J183434.9-084443                   & 4& $\sim 10$	   \\
 &             &            &          &  &         &       &   &                      	&        &B1830-08? &&3.5	   \\
 \hline
5&25.21 & -0.16 & 0.43 & 0.58&  1.e-5 & 15 & $6.4\pm 1.5$&  {\bf HESS J1837-069}&&J1838-0655&& 2.3	   \\
 \hline
6 &26.87 & -0.12 & 0.39 & 0.54 & 2.e-4 & 11 &$4.6\pm 1.4$& {\bf HESS J1841-055}&&J1841-0524 &4.9 &3.0				   \\
 \hline
7 &36.20 & 0.02   & 0.23 & 0.37 & 1.e-6 & 11 & $4.6\pm 1.4$& {\bf HESS J1857+026}&&J1856+0245 &10.3 & 2.0			   \\
\hline
 8 &78.09 & 2.54    &  0.33 & 0.38 &1.e-5 & 7&$2.3\pm 0.9$ &{\bf VER J2019+407}& $\gamma$Cyg&J2021+4026& &7.7\\
 \hline
      9&284.32 & -0.57& 0.32 & 0.42& 7.e-3 &  4 &$1.3\pm 0.7$&{\bf Westerlund 2} &&J1023-5746&&0.5		   \\
 \hline
10&287.12 & -0.80& 0.46 & 0.74& 2.e-4 & 9 &$2.9\pm 1.0$&  near Eta Car&&&&		   \\
 \hline
 11 &313.56  &0.11  & 0.2 &  0.32  &8.e-6&  8 &$2.6\pm 1.0$  & {\bf Kookaburra}       &        & J1420-6048  &7.7 &1.3			   \\
  \hline
12&331.66 & -0.58  & 0.27 & 0.64 & 7.e-4& 11 &$3.7\pm 1.1$ &{\bf HESS 1614-518}&&J1614-5144&&\\ 
 \hline
13&332.57 & -0.18  & 0.34 & 0.63 & 1.e-3 &10 &$3.3\pm 1.0$& {\bf HESS J1616-508}&&J1617-5055 &6.5 &0.8\\ 
 \hline
14&336.25 & 0.04 & 0.37 & 0.59 & 1.e-6 & 16  &  $5.4\pm 1.3$	&{\bf HESS J1632-478}&&J1632-4757 &7.0 &24   \\
 \hline
15&339.56  &-0.79 & 0.37 & 0.72&  3.e-3 &  10&$3.4\pm 1.0$& {\bf Westerlund 1} 	&&J1648-4611 &5.7 &11				   \\
 \hline
16       &344.90 & 0.23 & 0.72 & 1.05 & 3.e-2 & 8 & $2.8\pm 1.1$& {\bf HESS  J1702-420}&&J1702-4128?&5.2& 5.5 			   \\
 \hline
17       &346.20 & -0.31& 0.37 & 0.57 & 1.e-2 & 7 &$2.7\pm 1.0$&  {\bf HESS 1708-410}&&J1706-4009?&3.8&0.9			   \\
 \hline
18     &358.06 & -0.54& 0.57& 0.63& 1.e-4 &10 &$3.7\pm 1.2$&  {\bf HESS J1745-303}& &&	&			   \\
\hline
\end{tabular}
\caption{List of extended sources found in the MST analysis. $\theta_{50},\theta_{90}$ are the radii within which 50\% and 90\% of the tree photons are contained. $N_{ph}$ is the background subtracted number of photons in the tree.  $P_{90}$ is the chance coincidence probability of background fluctuation which would have $N_{ph}$ photons within a circle of the radius $\theta_{90}$. $F$ is the source flux in units of $10^{-11}$~erg/cm$^2$s. SNR and PSR columns give the names of supernova remnants and/or pulsars suggested to be associated to the source. $R_s$ and $T_s$ are the source age (in the units of $10^4$~yr) and distance (in kpc). }
\label{tab:list}
\end{table*}

The average radial brightness profile of the extended sources from Table \ref{tab:list} is shown in Fig. \ref{fig:psf}. One could see that the profile is clearly different from the LAT PSF. The cumulative photon distribution traces source photons up to $\theta\sim 0.6^\circ$.
The angular sizes of the sources reported in the Table \ref{tab:list} are generally larger than the sizes reported in the HESS survey \cite{aharonian06}. This difference might be attributed to the difference of the observation techniques used by the LAT and HESS telescopes.

\section{Details for individual extended  Galactic VHE \gr\ sources}

The estimates of the section \ref{sec:extended} show that recent sources of TeV CRs should be surrounded by degree-scale extended VHE \gr\ emission with the flux in excess of  $\sim 10^{-11}$~erg/(cm$^2$s). The sources listed in Table \ref{tab:list} satisfy this requirement. This means that they could trace recent injections of CRs in the Galaxy. In this section we investigate the nature of CR accelerators which might power the observed extended emission in each individual source.

\subsection{ $(l,b)=(8.15,-0.14)$}

\begin{figure}
\includegraphics[width=\columnwidth]{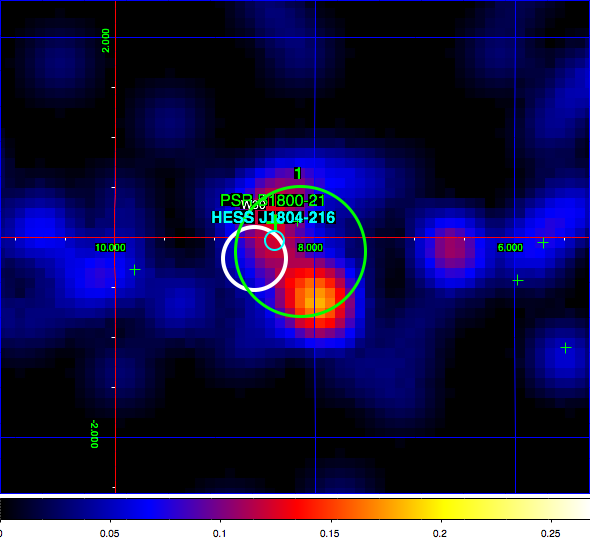}
\caption{Sky region around extended source at $(l,b)=8.15,-0.14)$. White circle marks the position of 2FGL extended source associated to supernova remnant W30. }
\label{fig:1804}
\end{figure}

This excess is associated to  HESS source HESS J1804-216, see Fig. \ref{fig:1804}. The size of the LAT event cluster associated to the source is clearly larger than the extent of the source measured by HESS.
The original discovery paper by HESS \cite{aharonian06} suggests a PWN interpretation of the source associating it to a Vela-like pulsar PSR J1800-21 \cite{kargaltsev07,rosat} with spin-down age $T_s\simeq 1.7\times 10^4$~yr situated at the distance $D\simeq 3-4$~kpc \cite{psrcat}. 

Alternatively, an association with a shell-type SNR could be considered. The SNR in question is G8.7-0.1, which is a relatively young ($10^4$~yr) and nearby  SNR visible in the 0.1-100~GeV energy band \cite{hanabata11}. Possible association of G8.7-0.1 with PSR J1800-21 is considered (see e.g. \cite{rosat}) based on the proximity of the two objects in space and similar age.

\subsection{$(l,b)=(17.58,-0.14)$ and  $(l,b)=(16.74,-1.79)$ }

\begin{figure}
\includegraphics[width=\columnwidth]{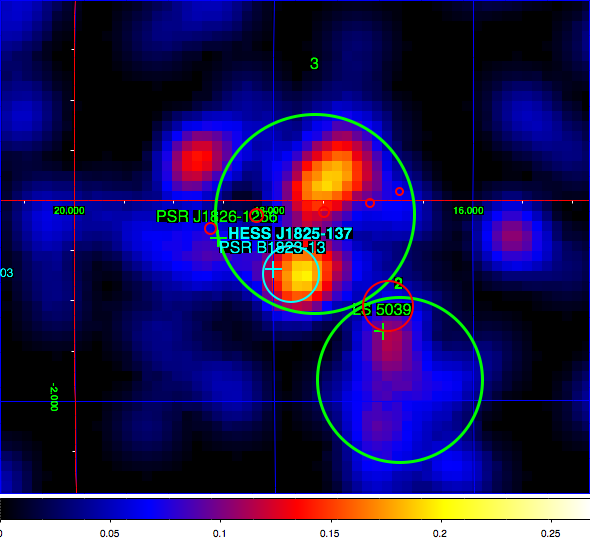}
\caption{Sky region around extended source at $(l,b)=17.58, -0.14)$.  Red circles show positions and sizes of SNRs from the Green catalogue \cite{green09}.}
\label{fig:1825}
\end{figure}

This extended emission region includes the HESS source HESS J1825-137, which is identified with the pulsar PSR B1823-13 \cite{aharonian06a,grondin11}, see Fig. \ref{fig:1825}. Similarly to PSR J1800-21, this is a relatively nearby ($4$~kpc) and young ($20$~kyr) pulsar \cite{psrcat}. 

The extended radio source, possibly a supernova remnant,  G 17.4-0.1 \cite{brogan}, is found within the source extension, so that relation of the extended emission to a shell-like SNR could not be ruled out. Lack of basic information on G 17.4-0.1 (age, distance) makes the analysis of association of the observed extended emission  to G~17.4-0.1 impossible.

Immediately adjacent to the extended source around HESS J1825-137, there is an extended emission which includes a \gr\ loud binary LS 5039 (Fig. \ref{fig:1825}). It is not clear if this source is just a further extension of a brighter HESS J1825-137, or it is an independent source. The relation of this source to LS 5039 is also not evident.

Energy-dependent morphology of HESS J1825-137 was studied in detail in the Ref. \cite{aharonian06a}. Detailed observations by HESS have revealed a more compact emission region of the size $\sim 0.3^\circ$ at the highest energies on top of a more extended degree-scale emission at the energies of several hundred GeV. Suzaku observaitons of the HESS J1825-137 region reveal an extended X-ray emission associated to the central more compact $0.3^\circ$ source, but not to the extended degree-scale source \cite{uchiyama08}.  The X-ray emission detected by Suzaku could be consistently interpreted as synchrotron emission from electrons in the PWN of PSR B1823-13. The absence of X-ray emission from the degree-scale source might indicate that the degree-scale \gr\ emission is due to CR interactions. 

\subsection{$(l,b)=(23.32, -0.16)$, $(25.21, -0.16)$ and $(26.87, -0.12)$.}

\begin{figure}
\includegraphics[width=\columnwidth]{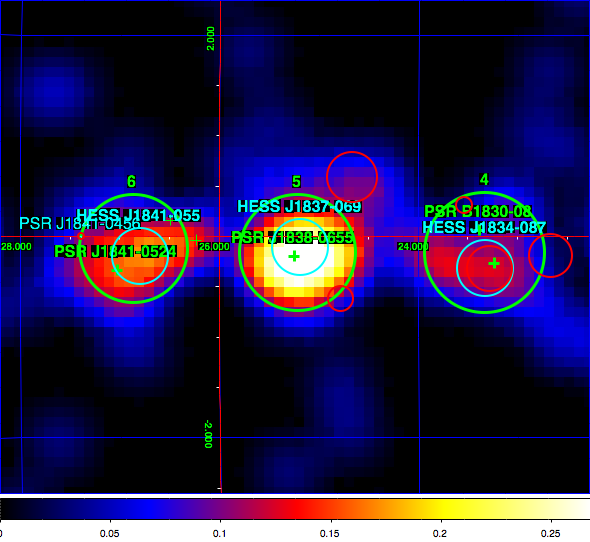}
\caption{Sky region around extended sources at $(l,b)=(23.32, -0.16)$, $(25.21, -0.16)$ and $(26.87, -0.12)$. Green cross inside HESS J1834-087 marks the position of CXOU J183434.9-084443. Red circles show positions and sizes of SNRs from the Green catalogue \cite{green09}.}
\label{fig:1837}
\end{figure}

The event clusters associated to these three adjacent excesses merge into one extended emission region along the Galactic Plane, in which the central source, the cluster at $(l,b)=(25.21 -0.16)$, dominates. This central source is associated to the HESS source HESS J1837-069 identified with a PWN of the pulsar PSR J1838-0655 \cite{gotthelf08,kargaltsev11}, which has the spin-down age of $\sim 22$~kyr \cite{psrcat}.

The  side sources are also associated to the HESS sources HESS J1834-087 and HESS J1841-055 (Fig. \ref{fig:1837}). The most probable low-energy counterpart of HESS J1834-087 is the SNR W41, which contains a central PWN-like object CXOU J183434.9-084443 \cite{misanovich11,mukherjee09}. Ref. \cite{tian07} provides an estimate of the age of this SNR, $T_s\sim 10$~kyr. The extension of the cluster of events associated to HESS J1834-087 is, however, larger than the size of the HESS source. Within this extension another relatively young pulsar is present, PSR B1830-08 (see Fig.\ref{fig:1837}), which somewhat younger, $T_s\simeq 30$~kyr.

The source associated to the HESS J1841-055 does not have a shell-like SNR candidates within its extension. At the same time, a young nearby pulsar PSR J1841-0524 of spin-down age $T_s\simeq 30$~kyr \cite{psrcat} is situated in the center of the source. This pulsar could be considered as a possible low energy counterpart of the extended \gr\ source. 

\subsection{$(l,b)=(36.20, 0.02)$}

\begin{figure}
\includegraphics[width=\columnwidth]{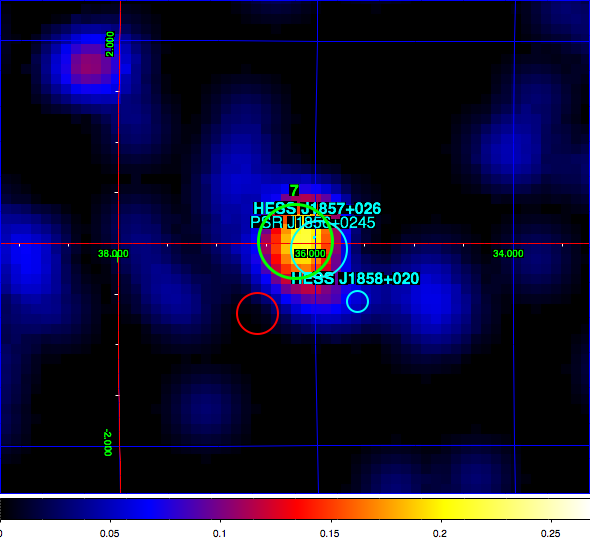}
\caption{Sky region around extended source at $(l,b)=(36.20, 0.02)$. Red circle shows position and size of a SNR from the Green catalogue \cite{green09}. }
\label{fig:1857}
\end{figure}

This extended source is spatially coincident with HESS source HESS J1857+026 and is identified with a PWN of PSR J1856+0245 \cite{hessels08,klepser11} of spin-down age $T_s\simeq 20$~kyr. No shell-like supernova remnant from the Green catalogue \cite{green09} is found within the source extent (see Fig. \ref{fig:1857}). The size of event cluster associated with the source is much larger than the reported size of the HESS source (see Fig. \ref{fig:1857}). The source is elongated in the direction of another HESS source, HESS J1858+020.

\subsection{$(l,b)=(78.09, 2.54)$}
\label{sec:cyg}

\begin{figure}
\includegraphics[width=\columnwidth]{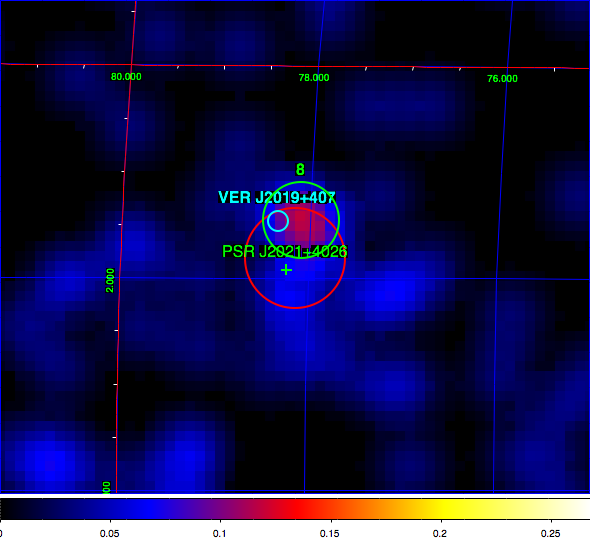}
\caption{Sky region around extended source at $(l,b)=(78.09, 2.54)$. Red circle shows the extent of $\gamma$Cygni SNR.}
\label{fig:2021}
\end{figure}

This source is a part of extended emission from the PSR J2021+4026 -- $\gamma$Cygni PWN+SNR system. The centroid of extended emission is displaced from the pulsar position, but the emission still comes from within the SNR boundary (Fig. \ref{fig:2021}), so that it is not the extended emission from the CRs escaping into the ISM. Taking into account the age of the SNR, $T_s\simeq 80$~kyr, and its proximity $R_s\simeq 1.5$~kpc \cite{landecker80} one could find that the expected size of the extended emission from the escaping CRs is $\theta_s\simeq  5^\circ$ (\ref{thetas}), i.e. it is expected to cover the entire Cygnus region. 

GeV band extended emission on the angular scale $\sim 5^\circ$ from the sky region immediately adjacent to the $\gamma$Cygni source was recently discussed in the Ref. \cite{ackermann}. This reference attributes the extended emission to the interactions of CRs in the ISM. Although the possibility of relation of this extended emission to $\gamma$Cygni is mentioned in this reference, an alternative possibility for the origin of CRs interacting in the ISM is considered, based on the spatial displacement of the source from the $\gamma$Cygni position.  The scenario favored in the Ref. \cite{ackermann} is injection of cosmic rays via acceleration at multiple shocks formed by stellar wind bubbles in the OB associations in the Cygnus region, which form the Cygnus superbubble.

We notice that in principle the displacement of the extended emission from the possible primary source position (pulsar) is expected, because of the presence of the ordered component of Galactic magnetic field. Ordered magnetic field introduces anisotropy in the CR diffusion from the source, so that the extended emission produced by CRs escaping from the source is not expected to be symmetrically distributed around the source position. Additional asymmetry could be introduced by spatial variations of the ISM density. 

It is also worth noticing that, contrary to the low-energy (GeV) emission,  the maximum of the extended emission in the VHE band is spatially coincident with the $\gamma$Cygni supernova remnant associated to PSR J2021+4026. This suggests that the higher energy (TeV) and lower energy (1-10~GeV) might be produced at different locations, via different acceleration mechanisms.

\subsection{$(l,b)=(284.32, -0.57)$ and near Eta Car}

\begin{figure}
\includegraphics[width=\columnwidth]{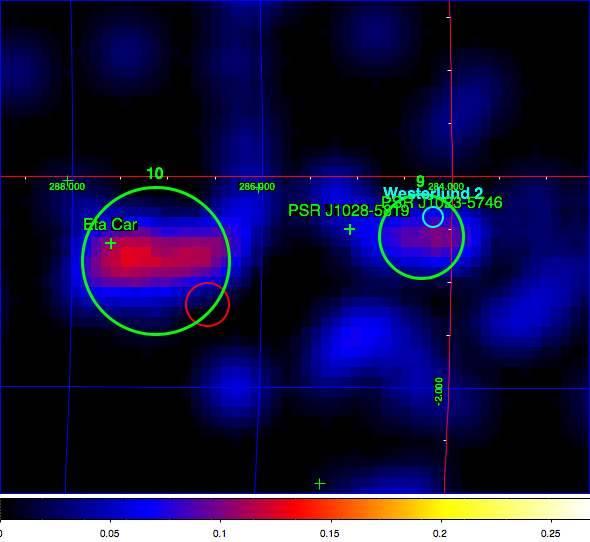}
\caption{Sky region around extended source at $(l,b)==(284.32, -0.57)$ and near Eta Car. Red circle shows position and size of a SNR from the Green catalogue \cite{green09}.}
\label{fig:etacar}
\end{figure}

The excess at $(l,b)=(284.32 -0.57)$ is positionally coincident with the HESS source  HESS 1023-575 identified as a PWN of PSR J1023-5746 (previously associated with Westerlund 2 cluster) \cite{abramowski11}. The pulsar  is relatively young, $T_s\simeq 5$~kyr. Absence of the radio emission does not allow the measurement of the distance based on the dispersion measure  \cite{psrcat}. No shell-like supernova remnant from the Green catalogue is visible within the source extension. 

No obvious counterpart candidates could be found within the extent of the emission region adjacent to the Eta Car (see Fig. \ref{fig:etacar}): no known pulsars or shell-like SNRs. The nature of this excess requires further investigation. The emission from this sky region comes from the direction tangential to the Carina-Sagittarius arm of the Galaxy (see Fig. \ref{fig:galaxy} below). The enhancement of the 100~GeV emission might, in principle, be an enhancement of Galactic diffuse emission  due to the projection effect. 

\subsection{$(l,b)=(313.56,0.11)$}

\begin{figure}
\includegraphics[width=\columnwidth]{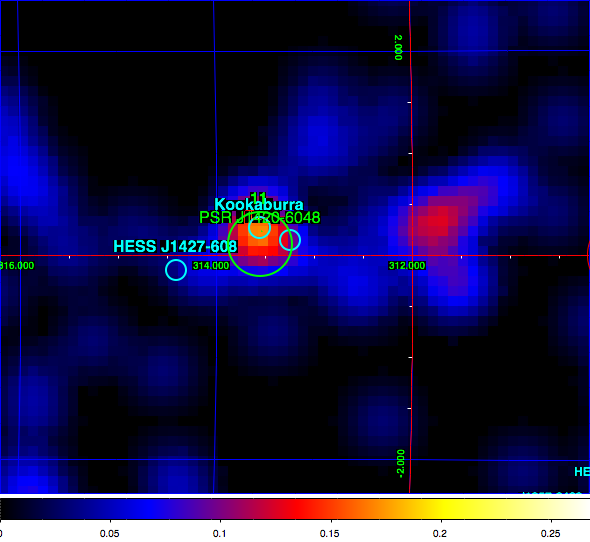}
\caption{Sky region around extended source around $(l,b)=(313.56,0.11)$. }
\label{fig:kooka}
\end{figure}

The extended source at $(l,b)=(313.56,0.11)$ is spatially coincident with the known HESS source at the Kookaburra PWN around PSR J1420-6048 of the age $13$~kyr situated at the distance 7.7~kpc \cite{psrcat}.

\subsection{$(l,b)=(331.66,-0.58)$, $(l,b)=(332.57,-0.18)$ and $(l,b)=(336.56,0.04)$  }

\begin{figure}
\includegraphics[width=\columnwidth]{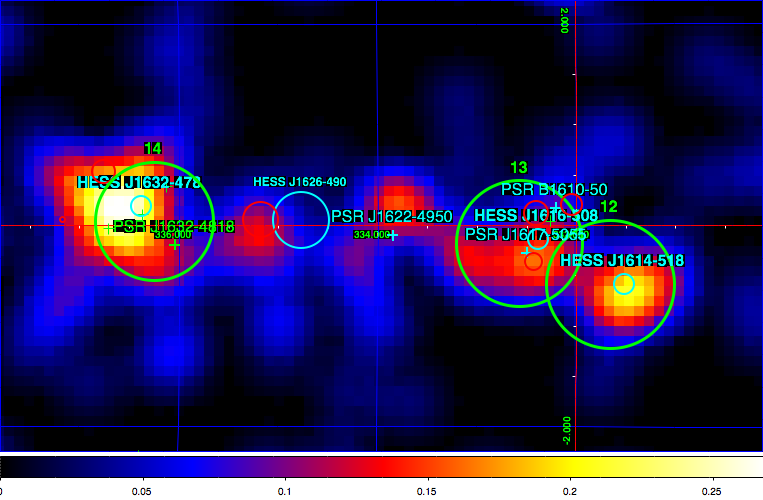}
\caption{Sky region around extended source around $(l,b)=(331.66,-0.58)$, $(l,b)=(332.57,-0.18)$, $(l,b)=(334.97,-0.26)$ and $(l,b)=(336.56,0.04)$. Red circles show positions and sizes of SNRs from the Green catalogue \cite{green09}.}
\label{fig:1614}
\end{figure}

The three excesses are close to each other so that the event clusters associated to these three sources merge into a single extended emission region along the Galactic Plane. The excesses could be identified with known HESS sources: HESS J1614-518, HESS J1616-508 and HESS J1632-4757 (see Fig. \ref{fig:1614}).  

The source at the position of HESS J1614-518 does not have any known pulsars or SNRs within its extent. The direction toward this source is tangent to the Norma arm of the Galaxy. Similarly to the case of the Scrutum-Centaurus arm, enhancement of the matter density in the arm and projection effects might be responsible for the enhanced VHE \gr\ flux from this direction.

HESS J1616-508 is most probably associated to the PWN of PSR J1617-5055 \cite{landi07} of the age $T_s\simeq 8$~kyr \cite{psrcat}, although several shell-like SNRs are adjacent to the HESS source position and are within the extent of the LAT event cluster at  $(l,b)=(332.57,-0.18)$ (see Fig. \ref{fig:1614}). 

The event cluster at the position of HESS J1632-478 is the brightest excess in this source group. The maximum of excess is at the position of the pulsar PSR J1632-4757 and Ref. \cite{balbo10} associates the source to the pulsar. In this case the age of the supernova associated to the source is $T_s\simeq 24$~kyr. Several other pulsars and SNRs, as well as another HESS source, HESS J1634-472, are found at the periphery of the source (see Fig. \ref{fig:1614}). It is possible that the overall emission from this bright excess is composed of contributions from several sources.

\subsection{Source at $(l,b)=(339.56,-0.79)$ }

\begin{figure}
\includegraphics[width=\columnwidth]{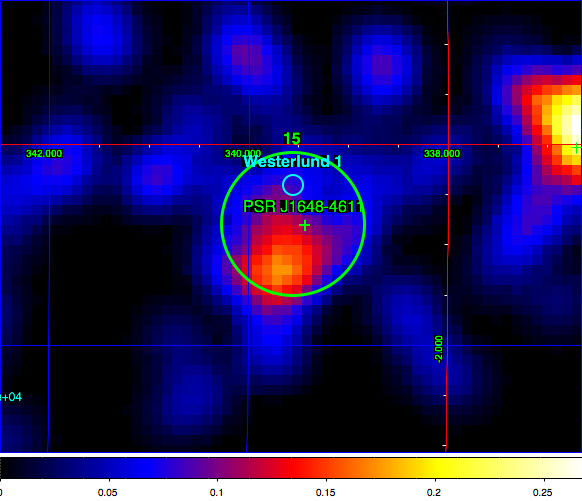}
\caption{Sky region around extended source at $(l,b)=(339.56,-0.79)$. }
\label{fig:1648}
\end{figure}

Excess at this position coincides with the HESS source in the direction of Westerlund 1 stellar cluster. The LAT event cluster associated to the source is more extended than the HESS source toward the position of the LAT pulsar PSR J1648-4611 (Fig. \ref{fig:1648}). This is a $T_s\simeq 100$~kyr pulsar situated in the Norma arm of the Galaxy at the distance $R_s\simeq 5$~kpc.

\subsection{$(l,b)=(344.90,0.23)$ and $(l,b)=(346.20,-0.31)$ }

\begin{figure}
\includegraphics[width=\columnwidth]{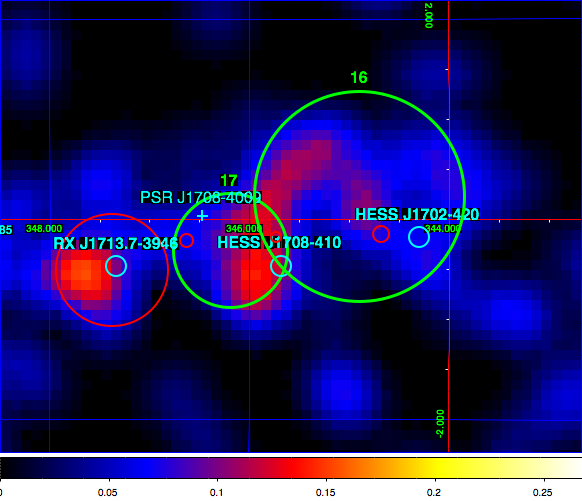}
\caption{Sky region around extended sources at $(l,b)=(344.90,0.23)$ and $(l,b)=(346.20,-0.31)$. Red circles show positions and sizes of SNRs from the Green catalogue \cite{green09}.}
\label{fig:1708}
\end{figure}

The LAT event cluster in this direction extends along the Galactic Plane between three HESS sources, HESS J1702-420, HESS J1708-410  (Fig. \ref{fig:1708}).  Several shell-like SNR and PWNe are found within the source extent, so that no "natural" counterpart candidates could be identified. Maximum of the excess at 100~GeV is at HESS J1708-410, but at slightly lower energies the maximum shifts toward HESS J1702-420, close to the pulsar PSR J1702-4128, so that the source morphology is energy-dependent. This pulsar has the age 55 kyr and is situated at the distance $\simeq 5$~kpc \cite{psrcat}. Another pulsar, PSR J1706-4009, is situated within the extent of the event cluster at $(l,b)=(346.20,-0.31)$. This pulsar is situated at the distance 3.8 kpc and has the spin-down age 9 kyr \cite{psrcat}. 

\subsection{Source at $(l,b)=(358.06,-0.54)$ }

\begin{figure}
\includegraphics[width=\columnwidth]{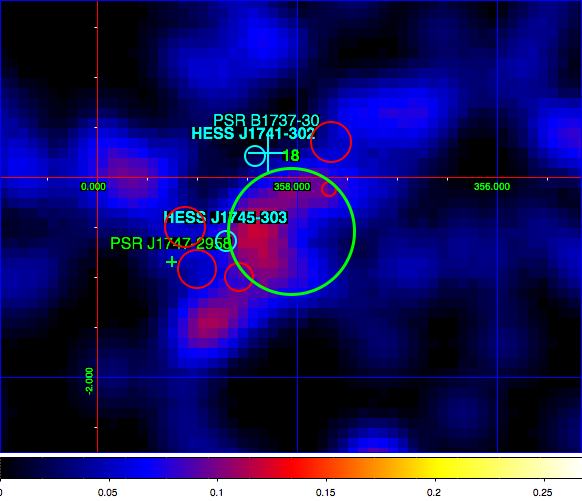}
\caption{Sky region around extended source at $(l,b)=(358.06,-0.54)$. Red circles show positions and sizes of SNRs from the Green catalogue \cite{green09}. }
\label{fig:1745}
\end{figure}

This excess is at the position of the HESS source HESS J1745-303, but has a much larger size than the HESS source. Several pulsars and shell-type SNRs are found within the source extension so that no obvious counterpart candidate could be singled out. 

\section{The nature of extended sources}

The overall number of extended 100~GeV \gr\ sources at the flux level in excess of $10^{-11}$~erg/cm$^2$~s is consistent with expected number of the regions of degree-scale extended emission around recent TeV CR sources \an{in scenario in which CRs are injected in portions of $\sim 10^{50}$~erg rather than continuously all over the Galaxy} (see  sections \ref{sec:locating}, \ref{sec:extended}).  The  source sizes are also consistent with an assumption that the extended emission is produced by high-energy particles diffusing into the ISM. Fifteen out of the eighteen  extended sources listed in the Table \ref{tab:list} are possibly or clearly associated with pulsars of the age $T_s\lesssim 30$~kyr. This is consistent with the hypothesis that TeV CRs are injected by supernova related phenomena. Indeed, majority of the core collapse supernova (more than 75\%) in the Galaxy are expected to result in production of the neutron star \cite{woosley}. Most of the collapsing massive stars are rotating, so that significant fraction of the new-born neutron stars are expected to be visible as pulsars \cite{heger05}.

\begin{figure}
\includegraphics[width=\columnwidth]{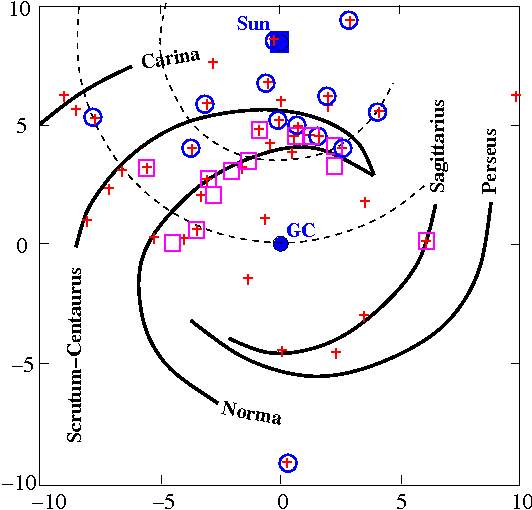}
\caption{Locations of the pulsars associated to the extended sources at 100~GeV on the Galaxy map.  Boxes mark the brighter excesses identified in the Fermi data, blue circles correspond to pulsars within weaker VHE \gr\ sources detected by the ground-based \gr\ telescopes. Red crosses mark the positions of known pulsars with ages in the range $T_s<3\times 10^4$~yr. Blue circle marks the position of the Galactic Center. Blue box is the position of the Sun. Solid curves show the locations of the main arms of the Galaxy. Dashed arcs show the distances 5~kpc and 8~kpc. }
\label{fig:galaxy}
\end{figure}

An essential element of the supernova scenario for the origin of CRs is that a significant part of supernovae (rather than just a small fraction of them) should inject some $10^{50}$~erg of CRs in the interstellar medium.  Assumption that every supernova injects $\sim 10^{50}$~erg in CRs implies that the 100 pc scale extended VHE \gr\ emission regions  should exist around nearly all the relatively recent (30~kyr) supernovae. Calculations of the previous sections suggest that extended emission regions situated at the distance $R_s\lesssim 5$~kpc should be detectable at the flux level in excess of $10^{-11}$~erg/cm$^2$s. 

The sensitivity of LAT in the $E>100$~GeV energy range is only marginally sufficient for detection of degree-size sources with fluxes $\sim 10^{-11}$~erg/cm$^2$s. In fact, all the sources listed in Table \ref{tab:list} have slightly higher fluxes. This means that, in principle, some extended emission regions around the nearby  TeV CR sources of the age $T\lesssim 30$~kyr might have escaped detection by LAT. At the same time, their flux level might be above the sensitivity of the ground-based \gr\ telescopes.  To verify this conjecture, we compiled a list of  all known pulsars with the spin-down age shorter than $30$~kyr using the ATNF pulsar catalogue \cite{psrcat}. As it is explained above, such pulsars could be used as tracers of recent supernova explosions. Fig. \ref{fig:galaxy} shows the locations of the pulsars in the Plane of the Galaxy. There are 56 such pulsars at the distances smaller than $20$~kpc. Some 29 of them are found within $0.5$ degrees from the known VHE \gr\ sources. The list the pulsars found in vicinity of known VHE sources, which are not detected in the MST analysis of LAT data is given in Table \ref{tab:hess}. In fact, a large fraction of the sources listed in Table \ref{tab:hess} are associated with 100 GeV \gr\ event clusters in LAT, but with the event statistics below the threshold adopted in our analysis.

The positions of the pulsars associated with the brightest extended VHE  \gr\ sources listed in Table \ref{tab:list} are marked by magenta boxes in Fig. \ref{fig:galaxy}. One could notice that most of the sources are situated in the Norma arm of the Galaxy, which is $\sim 5$~kpc away from the Sun. The fact that the extended sources in the Norma arm appear brighter than the extended emission around closer  CR sources might look surprising at the first sight. However, the Norma arm is a special region of the Galaxy with the highest star formation rate \cite{norma}. High star formation rate is associated to the increased matter density in the arm (presence of numerous molecular clouds), which boosts the VHE \gr\ luminosity of the extended emission. 
Positions of the pulsar-associated extended or point sources which are below the LAT sensitivity above 100~GeV, but possibly associated to the VHE \gr\ emission are marked by the blue circles in Fig. \ref{fig:galaxy}. 

From Fig. \ref{fig:galaxy} one can see that most of the known nearby supernovae associated with pulsars of the age $T_s\le 30$~kyr are spatially coincident with excess VHE \gr\ emission. In the case of sources listed in Table \ref{tab:list}, the emission is extended, with the typical source size in the degree range. This implies the spatial extent $\sim 100$~pc at the source distance.  Table \ref{tab:hess} includes both isolated (more compact) sources associated to the young pulsar wind nebulae (like e.g. MSH 15-52) and extended emission regions discussed above. Further investigation of the nature of emission from these sources is needed.  

\begin{table}
\begin{tabular}{l|l|l|l|l}
\hline
&Name & PSR & $R_s$& $T_s$\\
\hline
1& Vela X & B0833-45 & 0.29 &1.1\\
\hline
2&G292.2-0.5    &         J1119-6127 &8.40& 0.2\\
\hline
3&HESS J1303-631&  J1301-6305 &15.84&1.1 \\
\hline
4&HESS J1356-645 & J1357-6429,&4.09   &0.7\\
\hline
5&Rabbit                     &  J1418-6058 &         &1.0\\
\hline
6&MSH 15-52            &  B1509-58     &5.8 &0.2\\  
\hline
7&HESS J1708-443  & B1706-44     &1.82  &1.8  \\
\hline
8&HESS J1741-302  &  B1737-30     &3.28, &2.1\\ 
\hline
9&G0.9+0.1                &   J1747-2809 &$\ge 8$ &0.5\\
\hline
10&HESS J1809-193    & J1809-1943  &3.57  &1.1\\
    &                                    & J1811-1925  &    &2.3\\
\hline
11&HESS J1813-178  & J1813-1749  &          &0.5 \\
\hline
12&HESS J1833-105  & J1833-1034  & 4.30 &0.5\\
\hline
13&HESS J1846-029  & J1846-0258  & 5.10 &0.1\\
\hline
14&MGRO J1908+06  & J1907+0602 &3.01  &2.0\\
\hline
15&G54.1+0.3              & J1930+1852 &5.00  &0.3\\
\hline
16&MGRO J2019+37 & J2021+3651& 18.9& 1.7\\
\hline
17&Boomerang & J2229+6114& 3.0  &1.1\\
\hline
\end{tabular}
\caption{List of $T_s<30$~kyr pulsars positionally coincident with known VHE \gr\ sources}
\label{tab:hess}
\end{table}

\section{Discussion}

Existence of the 100~pc size extended VHE \gr\ emission regions around the locations of recent injections of TeV CRs in the Galaxy is an immediate testable prediction of the scenario of injection of CRs by the supernovae. These extended emission regions should be visible as degree scale VHE \gr\ sources with fluxes at the level of $\sim 10^{-11}$~erg/cm$^2$s above 100~GeV. In this paper we have searched for such degree-scale sources in the LAT data. We have found 18 such sources.  Table \ref{tab:list} gives the list of the extended sources found using the MST method. The overall extended source statistics, as well as the morphological properties of sources listed in Table \ref{tab:list}
agree with the theoretical expectations. Based on this agreement, we suggest that the bright extended sources are associated to the sites of recent CR injections in the Galaxy. 

The largest part ($80$\%) of the extended VHE \gr\ sources in the Galactic Plane, listed in Table \ref{tab:list} could be associated to pulsars. For eleven out of fifteen pulsar-related sources (i.e. 70\%) no shell-like SNR structure has been found, which might signify that the supernovae did not result in production of a regular expanding shell (similarly e.g. to the supernova which produced the Crab pulsar). Thus, it is not clear if the presence of a regular shell is an essential ingredient of the CR acceleration process. At the same time, all the shell-type SNRs entering Table \ref{tab:list} are composite SNRs confining a PWN.    None of the sources are associated to  shell-like SNR without a pulsar. This might indicate that  CR acceleration phenomenon is related to the presence of a pulsar in the SNR.

A significant fraction of pulsars is born in OB associations \cite{amnuel86,kaaret96}. Young pulsars which have typical birth kick velocities $\le 10^3$~km/s$=1$~pc/kyr  do not leave their parent association of the size $\sim 10-100$~pc until they reach the age $10^4-10^5$~yr. Taking this into account, it is possible that locations of the young pulsar in the Galactic Plane just trace the OB associations where they were born. In this case the degree-scale extended \gr\ emission might, in fact, be related to the presence of an OB association or a superbubble, which contains multiple supernova remnants and bubbles blown by the winds of massive stars.  The most clear example of this type is given by the extended source in the Cygnus region (source number 8 in Table \ref{tab:list}). Study of extended emission in the GeV band reported Ref. \cite{ackermann} finds that cosmic rays were recently injected in this region and suggests that the injection mechanism is related to the multiple shocks produced by the collective effect of winds from massive star filling the Cygnus region superbubble. An alternative possibility, also discussed in the Ref. \cite{ackermann} is that the cosmic rays are produced by the nearby $\gamma$Cygni supernova which hosts PSR J2021+4026 (see section \ref{sec:cyg}). 

Spatial coincidence of the extended VHE \gr\ sources in the Galactic Plane with the pulsars introduces a potential possibility that the extended emission is produced by electrons escaping from the PWN around the pulsar, rather than by CRs filling the 100~pc-scale bubble. This possibility requires further investigation. The VHE \gr\ emission in the $E\sim 100-300$~GeV range could be produced via inverse Compton scattering by electrons of comparable energies upscattering optical-infrared Galactic photon background (photons with energies $\sim 0.1-1$~eV). The inverse Compton cooling time of electrons is $\sim 2\times 10^6$~yr, assuming the density of the optical-infrared background $U\sim 1$~eV/cm$^3$, typical for the Norma Arm \cite{porter}. This cooling time scale is order-of-magnitude shorter than the CR cooling time (\ref{tpp}), calculated assuming the ISM density $n\sim 1$~cm$^{-3}$. This implies that electrons loose energy more efficiently than protons. To compare the luminosities generated by electrons to those produced by CRs, one needs to know the CR and electron densities in the relevant energy scales, which is highly uncertain. A first guess could be found from the ratio of electron and CR fluxes observed at the Earth: the electron flux at 300~GeV is $\sim 10^{-2}$ of the TeV CR flux \cite{fermi_electrons,pamela_cr}. Assuming this ratio, one finds that the CR contribution to the extended VHE source flux should dominate over the electron contribution.  It is, however, not clear if the locally observed electron-to-nuclei ratio could serve as a good estimate of the ratio close to the distant CR sources.  Electrons should still give a dominant contribution to emission from the central parts of the degree scale sources, at the locations of Crab-like PWNe around the pulsars. Two complementary measurements could be used to separate electron and CR contribution to the VHE \gr\ flux from the degree-scale sources. 

First, electron-dominated emission from the central PWN could be identified via radio-to-X-ray observations which could establish the morphology of the PWN and measure the spectrum of the synchrotron emission from high-energy electrons. At larger, degree scales, electrons with energies $E_e\sim 300$~GeV are expected to produce synchrotron emission in the $\epsilon_s\sim 10^{-2}\left[B/3\ \mu\mbox{G}\right]\left[E_e/300\mbox{ GeV}\right]^2$~eV energy range, i.e. in the far infrared (assuming magnetic field $B\sim 3\ \mu$G, typical for the ISM). Detection of the infrared counterparts of the extended VHE \gr\ sources will constrain the electron contribution to the degree-scale VHE source flux. 

Alternatively, proton/nuclei contribution to the extended source flux could be measured via detection of neutrino signal from the sources listed in Table \ref{tab:list} with km$^3$ scale neutrino detectors like IceCube \cite{icecube} and km3net \cite{km3net}. Asuming that most of the \gr\ emission from sources listed in Table \ref{tab:list} is produced by CR protons/nuclei, the neutrino flux is expected to be comparable to the \gr\ flux given in the column 8 of the Table. This would make the sources from Table \ref{tab:list} the strongest Galactic neutrino sources. The slope of the neutrino spectrum could be inferred from the HESS observations \cite{aharonian06}. The neutrino spectrum produced in nuclei-nuclei interactions is expected to follow the \gr\ spectrum, i.e. to have the powerlaw shape $dN_\nu/dE\sim E^{-2.4}$ in the energy range up to at least ten(s) of TeV. This slope is harder than the slope of the atmospheric neutrino spectrum, so that the sources should be preferably detectable at the highest energies. However, one should note that the source size is expected to increase with energy (see Eq. 5) making the sources extended for IceCube and km3net.  

IceCube is mostly sensitive to the sources in the Northern hemisphere. Only two of the 18 sources in Table \ref{tab:list} are situated in the Northern hemisphere: the source 8 in the Cygnus region and the source 9 associated to  HESS J1857+026. Currently existing upper limits on the point source fluxes with IceCube 40-string configuration \cite{ic40} at positions of these sources are still above the expected neutrino flux level, expected to be around  $\sim 10^{-11}$~erg/cm$^2$~s at the energy of $0.1$~TeV and still lower at the energies above 10~TeV at which IceCube is most sensitive.

Possible association of the degree-scale VHE \gr\ sources to pulsar/PWN systems, rather than to the shell-like SNRs without pulsars suggests a possibility that the presence of pulsar plays a significant role in the CR production. PWNe formed  by the collisions of pulsar winds with the SNR material  and/or with the interstellar medium are known to be very efficient particle accelerators.  In these systems most of the energy initially stored as the rotation energy of the neutron star is transferred to the high-energy particles. Most of our knowledge of particle acceleration mechanisms in PWNe is derived from the observations of radio-to-\gr\ emission from the accelerated electrons. In general, the standing relativistic shock at the interface of the pulsar wind and the supernova/interstellar medium should also accelerate protons and heavier nuclei. The energy stored in the rotation of the neutron star is 
\begin{equation}
E_{NS}=\frac{I\Omega_{ini}^2}{2}\simeq 3\times 10^{50}\left[\frac{P_{ini}}{10\mbox{ ms}}\right]^{-2}\mbox{ erg}
\end{equation}
 for a typical neutron star with the moment of inertia $I\simeq 1.5\times 10^{45}$~g~cm$^2$ and the angular velocity $\Omega_{ini}=2\pi/P_{ini}$ corresponding to the initial rotation period $P_{ini}$. If typical initial periods of the pulsars at birth are in the range $P_{ini}\lesssim 10$~ms, the rotation energy of the neutron star is sufficient for powering the CR production. 
 
Relativistic nature of the shock accelerating CRs could leave an imprint on the CR spectrum in the form of a hardening below the energy corresponding to the gamma factor of the shock \cite{lemoine}. Such a hardening might be responsible for the low-energy break observed in the spectrum of Galactic CRs \cite{prl}.

Periods of known pulsars  increase with time as a powerlaw $P\sim t^{1/(n-1)}$ where $n$ is the breaking index which, for most of the pulsars is in the range $2\lesssim n\lesssim 3$ \cite{lyne}. Observations of the powerlaw spin-down do not provide information on the initial rotation period of the neutron star at birth. The initial periods of rotation of neutron stars have to be estimated based on numerical modeling of stellar evolution, supernova explosions and of the subsequent evolution of the neutron stars. Numerical calculations of evolution of massive stars up to the point of gravitational collapse only now start to be able to account for the effects of the stellar rotation. The state-of-art simulations predict the initial period in the range of $P_{ini}\sim10$~ms for a typical star of 15 Solar masses, if a special mechanism for the removal of the angular momentum from the stellar core by strong magnetic field is assumed \cite{heger05}. Otherwise, the calculations suggest the initial periods in the range of $P_{ini}\sim 1$~ms in the absence of strong magnetic coupling of the core to the outer layers \cite{meynet11}. 

This implies that the rotation energy of the neutron star could, in principle, provide the power sufficient for production of CRs. If this case the presence or absence of a well-defined SNR shell is not of crucial importance for the acceleration. If the shell is present, the high-energy particles accelerated in the PWN would be confined by the magnetic fields in the shell, where they could interact, like it happens in the MSH 15-52 / PSR B1509-58 system. Otherwise, if the regular shell is not formed (like in the Crab nebula), particles would escape from the PWN directly into the interstellar medium. 
Alternatively, pulsars might play an "auxiliary" role in the CR acceleration process providing an initial injection of "pre-accelerated" particles in the SNR shell. Understanding of the role of pulsars in the CR acceleration process requires further investigation.

Analysis of abundances of elements in the cosmic ray flux suggests that the bulk of Galactic cosmic rays with energies below 10~GeV is accelerated in the environments containing a mix of supernova ejecta and older interstellar material typically present in OB associations and superbubbles \cite{lingenfelter07,binns08}. Moreover, the the time scale of acceleration is most probably longer than 100~kyr, an observation based on the observation of abundances of specific elements \citep{wiedebeck99}. This implies that low-energy cosmic rays are not produced out of the single supernova ejecta soon after the supernova explosion, but rather originate from the slow acceleration on the system of multiple shocks present in OB associations and superbubbles. Recent Fermi observations of \gr\ emission from the interactions of freshly accelerated cosmic rays in the Cygnus superbubble further support this  scenario.

At the same time, recent observations of several breaks in the interstellar  cosmic ray spectrum: softening above $\sim 10$~GeV \cite{prl} and hardening above 200~GeV \cite{pamela_cr} indicate the presence of multiple components in the cosmic ray spectrum. The component which dominates the low-energy cosmic ray flux and peaks in the 1-10~GeV range has soft spectrum above 10~GeV and becomes subdominant above 200~GeV. It is not clear if conclusion about  possible OB associations and/or superbubble origin of the dominant low-energy component holds also at the energies above 200~GeV, or a different acceleration mechanism is relevant at higher energies. As it is mentioned above, association of the degree-scale extended sources with young pulsar might reflect possible OB association membership to the pulsar parent stars, so that multiple acceleration mechanisms related to the pulsar wind nebulae, supernova remnants and to the entire star-forming region around the supernova/pulsar  might be responsible for the production of cosmic rays filling the 100~pc-size regions visible as the desgree-scale \gr\ sources.

\section{Conclusions.}

We have considered a possibility of localization of the sources of TeV CRs via detection of degree-scale extended VHE  \gr\ emission produced by CRs escaping into the ISM. We have identified extended sources of $E>100$~GeV \gr s in the Galactic Plane using the data of LAT and studied their statistics and morphological properties. The list of the degree-scale extended source is given in Table \ref{tab:list}. The average surface brightness profile of the extended sources is shown in Fig. \ref{fig:psf}.

The observed extended source properties suggest their interpretation as extended emission produced by CRs escaping from relatively young sources of TeV CRs. Most of the degree-scale extended sources are associated to nearby pulsars with the spin-down age $T\lesssim 30$~kyr, as expected in the model of supernova origin of  the TeV CRs (if pulsars are considered as tracers of the recent supernova events) {\bf and their star-forming regions}. We find that nearly all known young nearby pulsars are associated to excess extended VHE \gr\ emission. At the same time, none of the detected degree-scale VHE sources is associated to a SNR without a pulsar. 

The observed degree-scale extended emission could be produced by both the proton/nuclear CRs and electrons/positrons escaping from the sources. It is difficult to distinguish between the two contributions based on the \gr\ data only. We suggest that the two contributions could be distinguished based on the complementary observations in the infrared band and in the VHE neutrino channel. 

\section*{Acknowledgements}

We would like to thank F.Aharonian, T.Courvoisier,  M.Kachelriess and A.Taylor for fruitful discussions of the subject and useful comments on the manuscript text.  The work of AN is supported by the Swiss National Science Foundation grant PP00P2\_123426/1.


\end{document}